\documentclass{article}
\usepackage[letterpaper, margin=1in]{geometry}
\usepackage[dvipsnames]{xcolor}
\usepackage{amsmath, amssymb, amsthm, mathtools, bbm, caption, subcaption, tabularx, hyperref, pdfpages}
\usepackage{url, multirow, booktabs, soul}
\usepackage{hyperref}

\newcommand*{\abs}[1]{{\left|{#1}\right|}}

\newcommand*{\vc}[1]{\mathbf{#1}}

\DeclareMathOperator*{\sgn}{sgn}

\newcommand{\new}[1]{{\em #1\/}}            

\newcommand{\ourmethod}{WExT{}}
\newcommand{\SR}{\Omega_{\text{R}}}
\newcommand{\SRC}{\Omega_{\text{RC}}}
\newcommand{\PR}{\Phi_{\text{R}}}
\newcommand{\PRC}{\Phi_{\text{RC}}}
\newcommand{\PWR}{\Phi_{\text{WR}}}

\newcommand*{\citep}[1]{\cite{#1}}
\newcommand*{\citet}[1]{\cite{#1}}

\begin{document}
\title{A Weighted Exact Test for\\Mutually Exclusive Mutations in Cancer}
\author{Mark D.M. Leiserson, Matthew A. Reyna, Benjamin J. Raphael\footnote{To whom correspondence should be addressed.} \\
Department of Computer Science and\\ Center for Computational Molecular Biology\\
Brown University\\
Providence, 02912, USA}

\maketitle

\abstract{
\noindent \textbf{Motivation:}
The somatic mutations in the pathways that drive cancer development tend to be mutually exclusive across tumors, providing a signal for distinguishing driver mutations from a larger number of random passenger mutations. This mutual exclusivity signal can be confounded by high and highly variable mutation rates across a cohort of samples. Current statistical tests for exclusivity that incorporate both per-gene and per-sample mutational frequencies are computationally expensive and have limited precision.\\
\textbf{Results:}
We formulate a weighted exact test for assessing the significance of mutational exclusivity in an arbitrary number of mutational events. Our test conditions on the number of samples with a mutation as well as per-event, per-sample mutation probabilities. We provide a recursive formula to compute $p$-values for the weighted test exactly as well as a highly accurate and efficient saddlepoint approximation of the test. We use our test to approximate a commonly used permutation test for exclusivity that conditions on per-event, per-sample mutation frequencies. However, our test is more efficient and it recovers more significant results than the permutation test. We use our {\ul{W}eighted \ul{Ex}clusivity \ul{T}est (\ourmethod{}) software} to analyze hundreds of colorectal and endometrial samples from The Cancer Genome Atlas, which are two cancer types that often have extremely high mutation rates. On both cancer types, the weighted test identifies sets of mutually exclusive mutations in cancer genes with fewer false positives than earlier approaches.\\
\textbf{Availability:}
See \url{http://compbio.cs.brown.edu/projects/wext} for software.\\
\textbf{Contact:}
\href{mailto:braphael@cs.brown.edu}{braphael@cs.brown.edu}\\

\section{Introduction}

A key challenge in cancer genomics is distinguishing the small number of somatic mutations that drive cancer from the vast majority of mutations that accumulate randomly. The ability to distinguish these \emph{driver} mutations from the random \emph{passenger} mutations may lead to better understanding of cancer biology and personalized therapies customized to a tumor's mutational profile.  However, large scale cancer sequencing efforts such as The Cancer Genome Atlas (TCGA)~\citep{TCGA_COADREAD,TCGA_UCEC, TCGA_THCA} and the International Cancer Genome Consortium (ICGC) have shown that many driver mutations are rare across patient cohorts and thus distinguishing the driver mutations from the passengers by their frequency of occurrence is a difficult problem.

Driver mutations are hypothesized to group into a small number of pathways or hallmarks~\citep{Hanahan2011}, and this hypothesis is a widely accepted explanation for the observed mutational heterogeneity of cancer~\citep{Vogelstein2013}. Thus, researchers have developed methods to identify combinations of mutations using varying levels of prior knowledge, from pathway databases~\citep{Wendl2011, Mootha2003, Subramanian2005} to protein-protein interaction networks~\citep{Vandin2011, Ciriello2012, Ruffalo2015, Leiserson2015b}.

Because prior knowledge of pathways and interactions is often noisy or unavailable, \textit{de novo} methods that do not use prior information are advantageous. The vast number of possible combinations of mutated genes makes complete de novo discovery of combinations computationally and statistically intractable.  However, a number of methods~\citep{Miller2011, Vandin2012, Ciriello2012, Leiserson2013, Szczurek2014, Constantinescu2015, Babur2015, Leiserson2015, Kim2015, Kim2016} use the observation that mutations within the same pathway are often mutually exclusive across tumors~\citep{Thomas2007, Yeang2008}. These methods differ in how they score mutual exclusivity and in how they identify the best scoring set(s) of mutations.

The first type of score for mutual exclusivity is a combinatorial score, such as the scores employed in \citet{Miller2011, Vandin2012, Leiserson2013}. For example, in the Dendrix algorithm \citep{Vandin2012}, the score for a set $M$ of mutational events is the difference between the number of samples with a mutation in $M$ (coverage) and the number of mutations in $M$ occurring in more than one sample (coverage overlap). The advantage of a combinatorial score is that it is easy to compute, but it was observed by~\citet{Leiserson2015} and others that the score is often biased towards sets with frequently mutated genes.

The second type of score for mutual exclusivity is a statistical score~\citep{Ciriello2012, TCGA_AML, Szczurek2014, Constantinescu2015, Leiserson2015, Kim2015, Babur2015, Kim2016}. A particularly useful statistical score for exclusivity is based on the exact distribution that conditions on the observed number of mutated samples in each gene~\citep{Leiserson2015, Babur2015}. For a pair of mutations, such a test is a one-sided Fisher's exact test for independence~\citep{TCGA_AML, Leiserson2015, Babur2015}. For more than two genes, \citet{Leiserson2015} generalized the exact test to multi-dimensional contingency tables.  They introduced the CoMEt algorithm that computes a generalization of Fisher's exact test for event sets of any size using either an exact tail enumeration algorithm or an approximation. They showed that conditioning on the number of mutations in each events reduces bias towards frequently mutated events compared to combinatorial scores.

Statistical scores that condition only on mutation frequencies do not account for the variation in mutation rate among tumors. It has been observed that the number of mutations in a tumor can vary over several orders of magnitude (e.g., see~\citet{Vogelstein2013, Lawrence2013, Roberts2014}). For example, colorectal tumors with microsatellite stability have a median of 66 nonsynonymous mutations, but colorectal tumors with microsatellite instability have a median of 777 mutations~\citep{Vogelstein2013}. Another example is from \citet{TCGA_UCEC}, who classified a subset of TCGA endometrial cancers as ultramutated or hypermutated.

Another useful statistical test for mutual exclusivity conditions on \emph{both} the number of mutated samples in each event \emph{and} the number of mutated events in each sample~\citep{Ciriello2012, Kim2015}. Since computing this distribution exactly is not computationally efficient, permutation tests are used.  The permutation tests, which compare observed results to a number of samples (${\sim}10^4$) drawn from a null distribution, are more tractable than computing the $p$-value exactly on genome-scale data, but the significance of the score is directly limited by the number of permutations.  MEMo~\citep{Ciriello2012} computes the significance of the \emph{coverage} (number of mutated samples) of $M$ using this permutational distribution for sets of any size $k$. MEMCover~\citep{Kim2015} computes the significance of the exclusivity of pairs to search for exclusivity within, between, and across cancer types. Both MEMo and MEMCover restrict their analysis to sets of genes that interact in a protein-interaction network. WeSME~\citep{Kim2016}, which appeared while this paper was under review, computes the significance of exclusivity of pairs of genes with a less expensive approximation to the permutational distribution.  To our knowledge, there is no method for quickly computing the significance of mutual exclusivity conditioned on both the observed number of mutations per event and number of mutations per sample.

\subsection{Contributions}

We introduce a weighted test for mutual exclusivity that conditions on the frequency of each mutational event in a set $M$ and also incorporates the probability that each event is mutated in each sample. Our test was inspired by a model derived by~\citet{Manescu2015}, who computed the significance of the overlap between two sets of genes weights by gene length. We introduce the weighted exclusivity test to approximate the fixed gene and sample frequency permutation test quickly and accurately by estimating the mutation probabilities from the null distribution of the permutation test. We present a recursive formula for computing the $p$-value of this test exactly and derive a saddlepoint approximation for arbitrarily sized groups of genes. We show that the saddlepoint approximation is both a fast and accurate approximation of the permutational distribution.  We also demonstrate that the saddlepoint approximation can be used to rapidly compute the CoMEt statistical test, which is a special case of the weighted test where the mutation probabilities for a given gene are the same in each sample.  

We use our \ul{W}eighted \ul{Ex}clusivity \ul{T}est (\ourmethod{}) software to identify sets of exclusive mutations in hundreds of colorectal and endometrial cancers. Cancer of these types often have extremely high mutation rates (e.g., see \citet{Vogelstein2013}), which makes them difficult to analyze when conditioning only on the number of mutated samples per event. However, our weighted statistical test allows us to effectively condition on the number of mutated genes per sample, and we identify exclusive patterns of mutations in these cancers that were missed by earlier approaches. We find that the weighted test identifies more biologically interesting sets than CoMEt~\citep{Leiserson2015}. We expect that the weighted test for mutual exclusivity will prove useful for many cancer types where defects in DNA damage or environmental exposures, e.g., ultraviolet light, lead to very high mutation rates in some samples.

\section{Methods}
\label{sec:methods}

We introduce a new weighted test for mutual exclusivity that incorporates per-event, per-sample mutation probabilities, and we describe how to use particular instances of our test to approximate commonly used tests for mutual exclusivity, which we refer to as the \new{row exclusivity (R-exclusivity)} and \new{row-column exclusivity (RC-exclusivity)} tests.

First, in Section~\ref{sec:permutation-tests}, we describe the R-exclusivity and RC-exclusivity tests. Next, in Section~\ref{sec:weighted-exclusivity-test}, we introduce our new weighted test for mutual exclusivity, which we call the \new{weighted-row exclusivity (WR-exclusivity) test}, that incorporates event and sample mutation frequencies without using permutations. In Section~\ref{sec:approximating-permutation-tests}, we describe how to approximate the R-exclusivity and RC-exclusivity tests with the WR-exclusivity test. Then, in Section~\ref{sec:weighted-exclusivity-test-methods}, we provide a recursive formula for computing the WR-exclusivity $p$-value exactly, and we derive a fast and accurate saddlepoint approximation for the WR-exclusivity $p$-value. Finally, in Sections~\ref{sec:search-exclusivity}, we describe how we search for exclusive sets, and in Section~\ref{sec:implementation}, we describe our \ourmethod{} software.

We summarize the tests and our contributions in this paper in Table~\ref{table:tests}.

\begin{table*}
\centering
\footnotesize
\begin{tabular}{llll}
\toprule
Test                                    & $p$-value & Conditioning                                  & Algorithms \\ \midrule
Row (R) exclusivity                     & $\PR$     & Event frequencies                             & Tail enumeration (CoMEt), \\
                                        &           &                                               & \textbf{saddlepoint approximation}    \\
Row-column (RC) exclusivity             & $\PRC$    & Event and sample frequencies                  & Permutations\\
\textbf{Weighted-row (WR) exclusivity}  & $\PWR$    & Event frequencies and \textbf{per-event,}     & \textbf{Recursive formula}, \\
                                        &           & \textbf{per-sample weights ($W$)}             &  \textbf{saddlepoint approximation} \\
\bottomrule
\end{tabular}
\caption{Three tests for mutual exclusivity, the values that are fixed in each test, and different algorithms for computing the $p$-values associated with the tests. {\bf Boldface} entries indicate contributions by this manuscript.}
\label{table:tests}
\end{table*}

\subsection{Notation}
\label{sec:notation}

We observe the presence or absence of mutational events across a collection of samples. The presence of an event may reflect a variety of genomic (e.g., the canonical \emph{BRAF} V600E mutation or deletions in \emph{CDKN2A}), proteomic, and/or epigenomic alterations. In this work, we analyze single nucleotide variants and small insertion/deletions (indels) by gene. For the clarity of exposition, we will describe these events at the gene level, but our weighted test can accommodate a broader class of mutational events.

Let $\{g_i\}_{i=1}^m$ be a set of $m$ genes and $\{s_j\}_{j=1}^n$ be a set of $n$ samples. For each sample, we observe the presence of one or more mutations in each gene, and we record the presence or absence of mutations in a per-gene, per-sample binary mutation matrix $A \in \{0, 1\}^{m \times n}$, where $A = [a_{ij}]$ with $a_{ij} = 1$ if gene $g_i$ is mutated in sample $s_j$ and $a_{ij}=0$ otherwise.

Let $M \subseteq \{g_i\}_{i=1}^m$ be a set of $k$ genes.  The gene set $M$ has \new{co-occurring} mutations in sample $s_j$ if multiple genes are mutated in that sample, i.e., there exist distinct $g_i, g_\ell \in M$ such that $a_{ij} = 1$ and $a_{\ell j} = 1$. Alternatively, the gene set $M$ has a \new{mutually exclusive} mutation in sample $s_j$ if one and only one gene is mutated in that sample, i.e., there exists $g_\ell \in M$ such that $a_{\ell j} = 1$ and $a_{ij}=0$ for $g_i \in M \setminus \{g_\ell\}$. Our goal is to identify sets of genes with statistically significant numbers of mutually exclusive mutations.

Let $r_i = \sum_{j=1}^n a_{ij}$ be the number of samples with mutations in $g_i$, let $c_j = \sum_{i=1}^m a_{ij}$ be the number of genes with mutations in $s_j$, let $z_M$ be the number of samples with co-occurring mutations in $M$, and let $t_M$ be the number of samples with mutually exclusive mutations in $M$.

For any mutation matrix $B$, let $B(M)$ be the submatrix of $B$ with rows corresponding to the gene set $M$, and let $t_{B(M)}$ be the number of mutually exclusive mutations in $B(M)$. We will use $t_M = t_{A(M)}$.

\subsection{Permutation tests for mutual exclusivity}
\label{sec:permutation-tests}

We describe two different permutation tests for mutual exclusivity. First, the \new{row-exclusivity (R-exclusivity)} test finds the probability $\PR(M)$ of observing at least $t_M$ mutually exclusive mutations in a gene set $M$ given that each $g_i \in M$ is mutated in $r_i$ samples. We describe this test as the row-exclusivity test because it conditions on the row sums of the mutation matrix.

Formally, we define $\SR$ to be the set of mutation matrices with same row sums as $A$. Let $\mathcal{E}_\text{R} = \left\{ B \in \SR :\: t_{B(M)} \geq t_M \right\}$ be the set of mutation matrices with at least $t_M$ mutually exclusive mutations in $M$. Then
\begin{equation}
\label{eq:fixed-row-test}
\PR(M) = \frac{\abs{\mathcal{E}_\text{R}}}{\abs{\SR}}
\end{equation}
is the $p$-value of the R-exclusivity test.

Since the R-exclusivity test only conditions on the row sums of $A$, we can consider each row of $A$ independently. This implies that to compute $\PR(M)$, we use only the rows corresponding to $M$. Thus, for $k=2$, the $p$-value $\PR(M)$ is equal to the $p$-value from the one-sided Fisher's exact test, which computes the tail probability by summing the hypergeometric probability of $2 \times 2$ contingency tables with fixed margins. The hypergeometric probability of each contingency table is the proportion of matrices in $\SR$ that give a contingency table with those margins. Note also that, when $k = 2$, the probability of observing $t_M$ or more mutually exclusive mutations is equal to the probability of observing $z_M$ or more co-occurring mutations. \citet{Leiserson2015} generalized this test to $k > 2$ genes as part of the CoMEt algorithm.

The \new{row-column-exclusivity (RC-exclusivity)} test finds the probability $\PRC(M)$ of observing at least $t_M$ mutually exclusive mutations in a gene set $M$ given that each $g_i \in M$ is mutated in $r_i$ samples \emph{and} each $s_j$ is mutated in $c_j$ genes. We describe this test as the row-column-exclusivity test because it conditions on the row \emph{and} column sums of the mutation matrix.

Formally, we define $\SRC$ to be the set of mutation matrices with the same row and column sums as $A$. Let $\mathcal{E}_\text{RC} = \left\{ B \in \SRC :\: t_{B(M)} \geq t_M \right\}$ be the set of mutation matrices with at least $t_M$ mutually exclusive mutations in $M$. Then
\begin{equation}
\label{eq:fixed-row-column-distribution}
\PRC(M)
= \frac{\abs{\mathcal{E}_\text{RC}}}{\abs{\SRC}}
\end{equation}
is the $p$-value of the RC-exclusivity test. Since $\SRC$ depends on the row and column sums of $A$, we cannot consider the rows of $A$, or even $A(M)$, independently.

The RC-exclusivity test is related to the co-occurrence and mutual exclusivity tests used in \citet{Ciriello2012} and \citet{Kim2015, Kim2016} with a few key differences. First, \citet{Ciriello2012} use coverage (i.e., $t_M + z_M$) instead of exclusivity as the test statistic, while \citet{Kim2015, Kim2016} limit to pairs of genes.  Second, both~\citet{Ciriello2012} and \citet{Kim2015} use permutation tests that sample matrices from $\SRC$, so their $p$-values are limited by the number of draws (e.g., both use $10^4$ permutations).

\subsection{Weighted exact test for mutual exclusivity}
\label{sec:weighted-exclusivity-test}

We introduce a new weighted test for mutual exclusivity. The \new{weighted-row-exclusivity (WR-exclusivity)} test finds the probability $\PWR(M)$ of observing at least $t_M$ mutually exclusive mutations in a gene set $M$ given that $g_i \in M$ is mutated in $r_i$ samples \emph{and} a per-gene, per-sample mutation probability matrix $W$ that prescribes weights with the presence or absence of individual mutations. We describe this test as the weighted-row-exclusivity test because it conditions on the row sums of the mutation matrix and a probability weight matrix.

For our model, we assume that $\{X_{ij}\}_{j=1}^n$ is a set of mutually independent Bernoulli trials for each gene $g_i$ with success probabilities $W=[w_{ij}]$, i.e.,
\begin{equation}
\Pr(X_{ij}=\ell)
= \begin{cases} w_{ij}, & \text{if } \ell=1, \\ 1-w_{ij}, & \text{if } \ell=0, \end{cases}
\end{equation}
where $w_{ij}$ is the probability that gene $g_i$ is mutated in sample $s_j$.  Let $T_{M,j}$ be a random variable with $T_{M,j} = 1$ if $s_j$ has a mutually exclusive mutation in a gene set $M$ and $T_{M,j} = 0$ otherwise.  Therefore, $Y_i = \sum_{j=1}^n X_{ij}$ is a Poisson binomial distributed variable for the number of mutations in $g_i$ and $T_M = \sum_{j=1}^n T_{M,j}$ is a test statistic for mutual exclusivity indicating the number of mutually exclusive mutations in $M$.  We want to find the tail probability (commonly referred to as the $p$-value) of observing at least $t_M$ mutually exclusive mutations in $M$ given that $g_i$ is mutated in $r_i$ samples.  The WR-exclusivity $p$-value $\PWR(M)$ is the probability of observing at least $t_M$ mutations in a gene set $M$ under this model with
\begin{equation}
\label{eq:tail-prob}
\PWR(M)
= \Pr(T_M \geq  t_M\: |\: Y_M = \vc{r}_M)
\end{equation}
where $Y_M = [Y_i]_{i \in M}$ and $\vc{r}_M = [r_i]_{i \in M}$.

Note that, for any gene $g_i$, the assumption that $Y_i = r_i$ implies that
\begin{equation}
\label{eq:weight-matrix-row-sums}
\sum_{j=1}^n w_{ij} = \sum_{j=1}^n E[X_{ij}] = E\left[\sum_{j=1}^n X_{ij}\right] = E[Y_i] = r_i
\end{equation}
by the definitions of $\{X_{ij}\}_{j=1}^n$ and $Y_i$.

\subsection{Approximating the permutation tests with the weighted exclusivity test}
\label{sec:approximating-permutation-tests}

Each of the sets $\SR$ and $\SRC$ underlying the R-exclusivity and RC-exclusivity tests, respectively, determines a per-gene, per-sample weight matrix $W = [w_{ij}]$ by considering the probability $w_{ij}$ of observing a mutation in gene $g_i$ in sample $s_j$, i.e.,
\begin{equation}
\label{eq:probability-matrix}
W
= \frac{1}{\abs{\Omega}} \sum_{B \in \Omega} B
\end{equation}
where $\Omega \in \{\SR, \SRC\}$.
Since both $\SR$ and $\SRC$ fix the number of mutated samples per gene, the weight matrix $W$ in \eqref{eq:probability-matrix} with $\Omega \in \{\SR, \SRC\}$ satisfies \eqref{eq:weight-matrix-row-sums}. We define $W_R$ to be the weight matrix with $\Omega=\SR$ and $W_{RC}$ to be the weight matrix with $\Omega=\SRC$.

For the R-exclusivity test, each row of $B \in \SR$ can be considered separately, so \eqref{eq:probability-matrix} for the set $\SR$ is given by $W_R=[w_{ij}]$, with $w_{ij} = \frac{r_i}{n}$.

However, for the RC-exclusivity test, each row $B \in \SRC$ cannot be considered separately, so, to the best of our knowledge, there is no closed-form expression for \eqref{eq:probability-matrix} for the set $\SRC$. Therefore, we generate an empirical weight matrix $W_{RC}^N=[w_{ij}]$ for $\SRC$ by drawing $N$ matrices $\SRC^N$ uniformly at random from $\SRC$ and computing \eqref{eq:probability-matrix} with $\SRC^N$ instead of $\SRC$. We assume that there is a nonzero probability that any gene can be mutated in any sample, and thus set $w_{ij}=\frac{1}{2N}$ when no mutation in gene $g_i$ is observed in sample $s_j$ in $\SRC^N$.

Estimating $W^N_{RC}$ in this way gives an accurate approximation of $\PRC(M)$ using relatively small values of $N$.

\subsection{Computing the weighted exclusivity test}
\label{sec:weighted-exclusivity-test-methods}

Our weighted test for mutual exclusivity requires computing the tail probability in \eqref{eq:tail-prob}, which can be computationally expensive. We compute the tail probability using two different strategies: a recursive formula and a saddlepoint approximation.

\subsubsection{Recursive formula for the weighted exclusivity test}
\label{sec:weighted-exclusivity-test-recursive-formula}

We present a recursive formula for computing the tail probability in \eqref{eq:tail-prob} exactly for sets $M$ of any size $k$.
Assuming that $\{Y_i\}_{i=1}^m$ are mutually independent, we can write \eqref{eq:tail-prob} as
\begin{equation}
\label{eq:tail-prob-frac}
\PWR(M)
= \frac{\Pr(T_M \ge t_M, Y_M=\vc{r}_M)}{\prod_{i \in M}\Pr(Y_i=r_i)}.
\end{equation}
Without loss of generality, let $M = \{1, \dots, k\}$.
We first find the joint probability in the numerator of \eqref{eq:tail-prob-frac} using a recursive formula, where $\Pr(T_M \geq t_M, Y_M = \vc{r}_M) = F(t_M, r_1, \dots, r_k, n)$ is computed by the recurrence relation
\begin{equation}
\label{eq:joint-prob}
\begin{split}
    & F(t, x_1, \dots, x_k, j) = \\
    & \sum_{\pi \in \{0,1\}^k} \prod_{i=1}^k q_{ij\pi_{i}} F(w_\pi(t), y_{\pi_1}(x_1), \dots, y_{\pi_k}(x_k), j-1),
\end{split}
\end{equation}
where
\[
    q_{ij\ell} = \begin{cases} p_{ij} & \text{if $\ell = 1$,} \\ 0 & \text{otherwise,} \end{cases}
\]
\[
    w_{\pi}(t) = \begin{cases} t-1 & \text{if $\sum_{i=1}^k \pi_i = 1$,} \\ t & \text{otherwise,} \end{cases}
\]
and
\[
    y_\ell(x) = \begin{cases} x-1 & \text{if $\ell = 1$,} \\ x & \text{otherwise.} \end{cases}
\]
The base cases for \eqref{eq:joint-prob} are
\begin{equation}
	F(t, x_1, \dots, x_k, j) =
	\begin{cases}
	1, & \text{if } t = x_1 = \dots = x_k = j = 0, \\
	0, & \text{if } \min\{t, x_1, \dots, x_k, j\} < 0, \\
	   & \quad t > \sum_{i=1}^k x_i, \text{ or } \\
	   & \quad \max_{i=1}^k x_i > n. \\
	\end{cases}
\end{equation}
We then find the marginal probabilities in the denominator of \eqref{eq:tail-prob-frac} using dynamic programming, which is a standard method for computing the Poisson-Binomial probability mass function~\citep{Hong2013}.

\subsubsection{Saddlepoint approximation for the weighted exclusivity test}
\label{sec:weighted-exclusivity-test-saddlepoint-approximation}

We derive a saddlepoint approximation~\citep{Butler2007} for computing the tail probability in \eqref{eq:tail-prob}. This approach is inspired by~\citet{Manescu2015}, who derive a saddlepoint approximation for an enrichment test for differentially expressed genes in Gene Ontology categories weighted by gene lengths. The saddlepoint approximation is specifically designed to provide a quick and accurate approximation of the tail probability. We present the key equation in \eqref{eq:saddlepoint-approximation} and provide a fuller derivation for $k = 3$ in the supplement.  The saddlepoint approximation is given by
\begin{equation}
\label{eq:saddlepoint-approximation}
\Pr(T_M \geq  t_M\: |\: Y_M = \vc{r}_M)
\approx 1 - \Phi(\tilde w) - \phi(\tilde w)\left(\frac{1}{\tilde w}-\frac{1}{\tilde u}\right),
\end{equation}
where $\Phi$ and $\phi$ are, in this setting, the cumulative distribution and density functions, respectively, of the standard normal distribution, and $\tilde w$ and $\tilde u$ are defined as follows.

Without loss of generality, let $M = \{1, \dots, k\}$.
First, for $\lambda \in \mathbb{R}^{k+1}$, let $\mathcal{M}_{Y_M, T_M}(\lambda) = E[e^{\sum_{i \in M}\lambda_iY_i+\lambda_{k+1}T_M}]$ be the joint moment generating function of $\{Y_i\}_{i \in M}$ and $T_M$, and let $\mathcal{K}_{Y_M, T_M}(\lambda) = \log \mathcal{M}_{Y_M, T_M}(\lambda)$ be the corresponding joint cumulant generating function.
Similarly, let $\mathcal{M}_{Y_i}(\lambda) = E[e^{\lambda_iY_i}]$ be the moment generating function of $Y_i$, and let $\mathcal{K}_{Y_i}(\lambda) = \log \mathcal{M}_{Y_i}(\lambda)$ be the corresponding cumulant generating function.

Next, let $\mathcal{K}'_{Y_M, T_M}(\lambda)$ and $\mathcal{K}''_{Y_M, T_M}(\lambda)$ be the gradient vector and Hessian matrix, respectively, of $\mathcal{K}_{Y, T}(\lambda)$, and let $\mathcal{K}'_{Y_i}(\lambda)$ and $\mathcal{K}''_{Y_i}(\lambda)$ be the gradient vector and Hessian matrix, respectively, of $\mathcal{K}_{Y_i}(\lambda)$.

Finally, define $\tilde w$ by
\begin{equation}
\tilde w
= \sqrt{2} \sgn\left(\tilde y_{k+1}\right)
    \sqrt{\sum_{i\in M} \mathcal{K}_{Y_i}(\hat x_i) - \mathcal{K}_{Y_M, T_M}(\tilde y) - \tilde y^T\left(\hat x - \tilde x\right)}
\end{equation}
and $\tilde u$ by
\begin{equation}
\tilde u
= 2\sinh\left(\frac{\tilde y_{k+1}}{2}\right)
    \sqrt{\frac{\abs{\mathcal{K}''_{Y_M, T_M}(\tilde y)}}{\prod_{i \in M} \mathcal{K}_{Y_i}''(\hat x_i)}},
\end{equation}
where
$\tilde x = \left(r_1, \dots, r_k, t_M-\frac{1}{2}\right)$ and $\tilde y = \left(\tilde y_1, \dots, \tilde y_{k+1}\right)$ with $\tilde y$ the unique solution for $\mathcal{K}'_{Y_M, T_M}\left(\tilde y\right) = \tilde x$ and \eqref{eq:saddlepoint-approximation} undefined if $\tilde y_{k+1} = 0$,
and
$\hat x = \left(\hat x_1, \dots, \hat x_k, 0\right)$ with $\hat x_i$ the unique solution for $\mathcal{K}'_{Y_i}(\hat x_i) = r_i$.

\subsection{Searching for sets of mutually exclusive mutations}
\label{sec:search-exclusivity}

Our goal is to identify sets $M$ of genes with significantly exclusive mutations, i.e., extremely small $p$-values $\PWR(M)$. There has been a considerable amount of work on methods for optimizing scores for mutually exclusive mutations, including Markov chain Monte Carlo methods~\citep{Vandin2012, Leiserson2015}, integer linear programs~\citep{Leiserson2013, Zhang2014}, greedy algorithms~\citep{Babur2015}, and others. These methods have been shown to be able to search datasets of many hundreds of mutation events for mutually exclusive mutations. Many of these methods can be modified to use our weighted exclusivity test to identify the most significant sets.

Since our focus is a statistical test for exclusivity, we instead enumerate all sets $M$ of $k$ genes satisfying the following basic criteria and test them with the R-exclusivity, RC-exclusivity, and WR-exclusivity tests:
\begin{enumerate}
\item The number $t_M$ of samples with mutually exclusive mutations must be larger than the number $z_M$ of samples with co-occurring mutations, i.e., $t_M > z_M$.
\item Each gene $g_i \in M$ must have at least one exclusive mutation.
\end{enumerate}
We use the Benjamini-Hochberg procedure~\citep{Benjamini1995} to control the false discovery rate (FDR). We examine the subset of genes in each dataset with a minimum mutation frequency so that we can enumerate and test all combinations of genes of a certain size in a reasonable amount of time.

\subsection{Implementation}
\label{sec:implementation}

We implemented the recursive formula for the WR-exclusivity test in Python and C, and we implemented the saddlepoint approximation for the WR-exclusivity test in Python using the NumPy and SciPy numerical libraries. We implemented the RC-exclusivity test in Python, and we used a bipartite double edge swap algorithm (see \citep{Milo2003, Gobbi2014}) that has been shown empirically to sample uniformly from $\SRC$. Our code, along with commands and data for reproducing the results and figures in this paper, is available as the \new{\ul{W}eighted \ul{E}xclusivity \ul{T}est (WExT)} software package at \url{http://compbio.cs.brown.edu/projects/wext}.

\section{Results}
\label{sec:results}
We compare the results of the WR-exclusivity test to both the R-exclusivity and RC-exclusivity tests on real data. In general, we can choose any weights to compute WR-exclusivity, but, in this paper, we specifically consider weights to allow us to approximate the R-exclusivity and RC-exclusivity tests. We use \ourmethod{} to discover mutually exclusive sets of mutations in thyroid, colorectal, and endometrial cancers, restricting our analysis to mutations at the gene level.

The rest of this section is organized as follows. In Section~\ref{sec:data}, we describe the data used in our experiments. In Section~\ref{sec:unweighted}, we compare the tail enumeration and saddlepoint approximation algorithms for computing the R-exclusivity $p$-values $\PR(M)$, and we show that the saddlepoint approximation provides a fast and accurate approximation for $\PR(M)$. In Section~\ref{sec:weighted}, we compare the results of the recursive and saddlepoint approximation algorithms for computing the WR-exclusivity $p$-values $\PWR(M)$ with the results of the RC-exclusivity test, and we show that $\PWR(M)$ is an accurate approximation of $\PRC(M)$ using either the recursive or saddlepoint approximation algorithms. In Section~\ref{sec:approximation}, we show that $\PWR(M)$ provides an accurate approximation of $\PRC(M)$ even with coarser estimates of the weight matrix $W$. Finally, in Sections~\ref{sec:thca} and~\ref{sec:coadread-ucec}, we present the results of the WR-exclusivity test on thyroid, colorectal, and endometrial cancers.

\subsection{Data}
\label{sec:data}
We analyzed non-synonymous single nucleotide variants (SNVs) and small insertions or deletions (indels) in 224 colorectal (COADREAD)~\citep{TCGA_COADREAD}, 402 papillary thyroid carcinoma (THCA)~\citep{TCGA_THCA}, and 248 uterine corpus endometrial carcinoma (UCEC)~\citep{TCGA_UCEC} samples from The Cancer Genome Atlas (TCGA). We analyzed the mutations in the COADREAD and UCEC samples from the TCGA Pan-Cancer project~\citep{Weinstein2013} by downloading the mutations in Mutation Annotation Format (MAF) from Synapse~\citep{1}. We downloaded the mutations in THCA from Firehose~\citep{2}. We restricted our analysis to non-synonymous mutations, ignoring mutations classified as ``Silent'', ``Intron'', ``3'UTR'', ``5'UTR'', ``IGR'', ``lincRNA'', and ``RNA''. We also downloaded lists of hypermutator samples for COADREAD and UCEC. We created a list of 35 hypermutator samples in COADREAD listed in~\citet{TCGA_COADREAD} in their Supplementary Table 3, and 82 hypermutator samples in UCEC listed by~\citet{TCGA_UCEC} as samples labeled ``POLE OR MSI'' in their Supplementary Datafile S1.1. We restrict our analysis to genes mutated in at least 20, 5, and 30 samples in the COADREAD, THCA, and UCEC datasets, analyzing 76, 30, and 62 genes in each dataset, respectively.

In general, COADREAD samples have the most {mutated genes (median: {78.5), with COADREAD hypermutators with mutations in at least an order of magnitude more genes than non-hypermutators (median for hypermutators: 797; median for non-hypermutators: 69). THCA samples have the fewest mutated genes per sample (median: 12), with no hypermutators, while UCEC has more mutated genes per sample (median: 57.5) with UCEC hypermutators mutated in approximately an order of magnitude more genes than non-hypermutators (median hypermutators: {355; median non-hypermutators: 43.5). See Supplementary Figure S1.

For each dataset, we estimated the weights $W^N_{RC}$ using the permutation procedure described in Section~\ref{sec:approximating-permutation-tests} using $N = 10^3$ permutations. We show the weights for each dataset in Figure~\ref{fig:weight-matrices}.

\begin{figure*}[t]
  \centering
  \includegraphics[width=\textwidth]{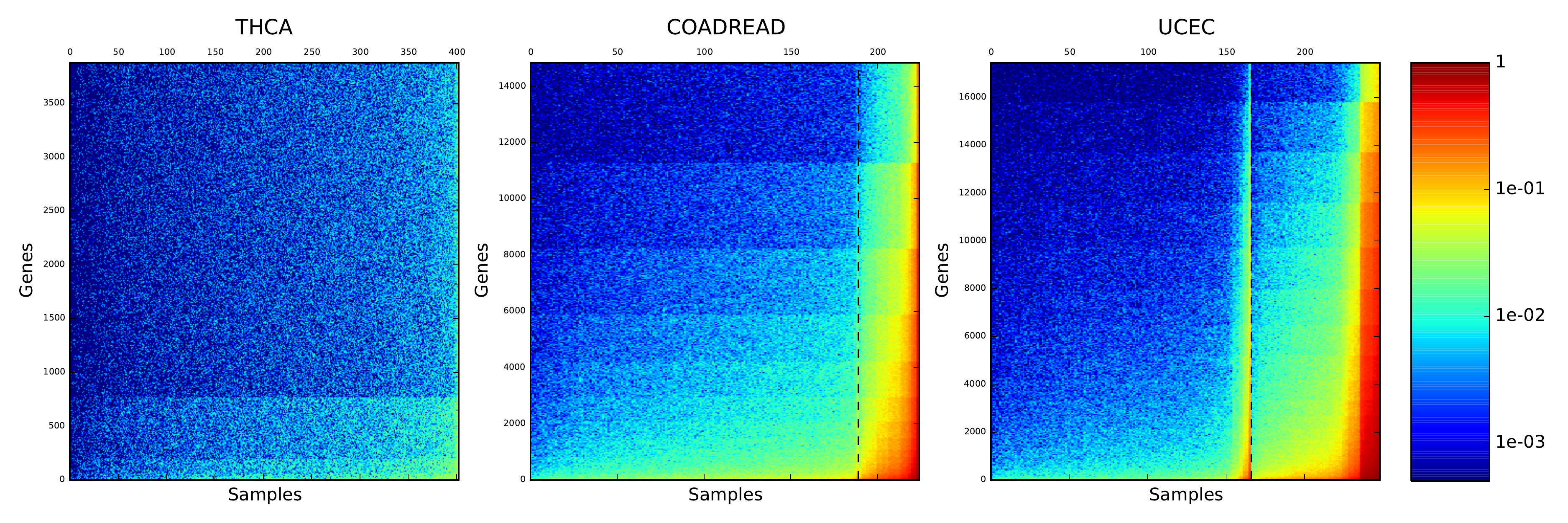}
  \caption{The weights $W_N$ estimated by sampling $N = 10^3$ permuted matrices on the THCA, COADREAD, and UCEC datasets. Samples ($x$-axis) are sorted by the number of mutated genes in increasing order from left to right, with hypermutators (right) separated from non-hypermutators (left) with a dashed line in COADREAD and UCEC. Genes ($y$-axis) are sorted by the number of mutated samples in increasing order from top to bottom.}
  \label{fig:weight-matrices}
\end{figure*}

\subsection{Comparison of methods for computing the R-exclusivity test on real data}
\label{sec:unweighted}
First, we investigated the accuracy and speed of the saddlepoint approximation of the R-exclusivity $p$-value $\PR(M)$. We enumerated triples according to the procedure described in Section~\ref{sec:search-exclusivity} in the THCA, COADREAD, and UCEC datasets, and computed $\PR(M)$ exactly using the CoMEt software from~\citet{Leiserson2015} as well as approximately using the saddlepoint approximation with $W_R$ given in Section~\ref{sec:approximating-permutation-tests}.

Supplementary Figure S2 shows a comparison of the $p$-values and runtimes given by the two methods, where the weights for the WR-exclusivity test are uniform across samples. On these datasets, the saddlepoint approximation is an extremely accurate approximation of the tail enumeration procedure ($\rho^2=0.995$). Additionally, while the median runtimes of the two algorithms are similar, the tail enumeration procedure is much slower for sets with co-occurring mutations while the saddlepoint approximation is largely unaffected.  We expect the discrepancy between runtimes to grow for gene sets of larger sizes.

\subsection{Comparison of methods for computing the WR-exclusivity test on real data}
\label{sec:weighted}
Next, we compared the results of methods for computing WR-exclusivity test with weights $W^N_{RC}$ the RC-exclusivity test on pairs of genes from the THCA, COADREAD, and UCEC datasets. We chose pairs instead of triples because of the prohibitive cost of computing the recursive formula for $\PWR(M)$. We used $N = 10^4$ permutations to compute the $\PRC(M)$, and also included the tail enumeration procedure for $\PR(M)$ as a control.

Table \ref{table:correlation-permutational} shows that the results of WR-exclusivity test -- computed either with the recursive formula or the saddlepoint approximation -- are strongly correlated with the RC-exclusivity test (Figure~\ref{fig:qqplots-and-runtime}(b)). The results of the R-exclusivity test are more weakly correlated with the RC-exclusivity test (Figure~\ref{fig:qqplots-and-runtime}(a)), showing that conditioning on the number of mutations in each sample changes the distribution of mutually exclusive mutations.  This discrepancy remains when we restrict to gene sets $M$ with $\PRC(M) \ge 10^{-4}$, i.e., sets of genes for which the empirical permutational distribution finds at least one mutually exclusive mutation in $M$.

The WR-exclusivity $p$-values computed exactly and with the saddlepoint approximation are highly correlated (Figure~\ref{fig:qqplots-and-runtime}(c)), with a Pearson's correlation coefficient of 0.996 for all $p$.  For smaller $p$-values with $\PWR(M) < 10^{-4}$ from either the recursive formula or the saddlepoint approximation, the correlation increases to 0.9999.

\begin{table*}
\centering
\begin{tabular}{lccc}
\toprule
\textbf{Pairs} & \textbf{$\PR$ (CoMEt)} & \textbf{$\PWR$ (recursive)} & \textbf{$\PWR$ (saddlepoint)} \\
\midrule
All & 0.71291 & 0.99816 & 0.99481 \\
$\PRC(M) \ge 10^{-4}$ & 0.65376 & 0.99811 & 0.99404 \\
\bottomrule
\end{tabular}
\caption{Pearson's correlation coefficient $\rho^2$ of $p$-values of pairs of genes from the THCA, COADREAD, and UCEC datasets using the tail enumeration R-exclusivity $p$-values $\PR(M)$, RC-exclusivity $p$-values $\PRC(M)$, and the recursive formula and saddlepoint approximations of the WR-exclusivity $p$-values $\PWR(M)$ using weights $W_R$. The correlations were computed for two sets of pairs of genes: $p$-values for 5,014 pairs (all) and 4,926 pairs ($\PRC(M) \ge 10^{-4}$).}
\label{table:correlation-permutational}
\end{table*}

The runtime to compute $\PWR$ using the recursive formula varies widely because pairs with co-occurring mutations require more computation, but the runtime of the saddlepoint approximation is more consistent.  As a result, testing all pairs with the recursive formula requires approximately 2 hours, but testing the same pairs with the saddlepoint approximation requires approximately 30 seconds. Note that the runtime does not include generating the weights $W^N_{RC}$, which requires several minutes.

\begin{figure*}[t]
  \centering
  \includegraphics[width=\textwidth]{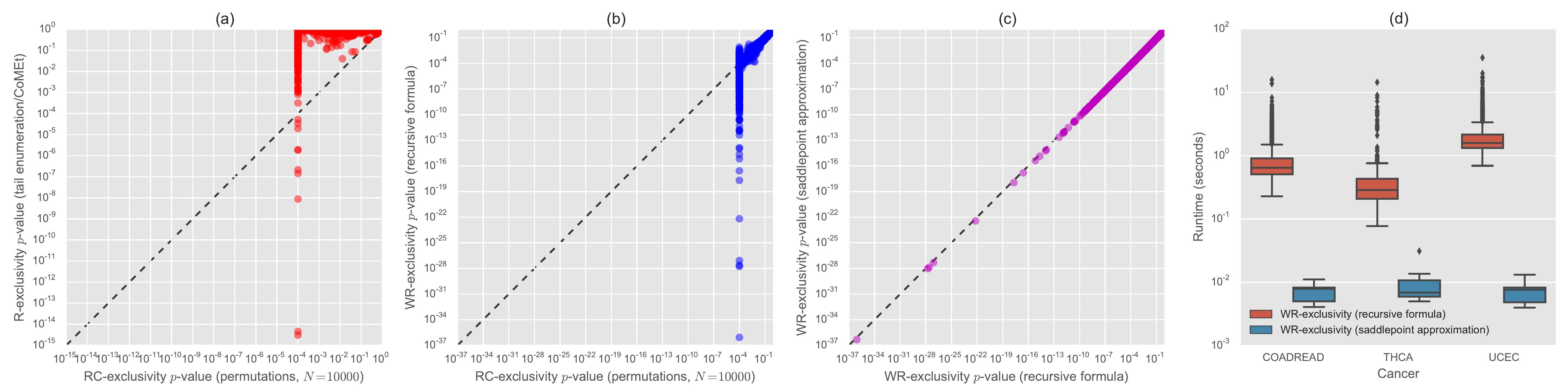}
  \caption{Comparison of $p$-values and runtimes of different tests on THCA, COADREAD, and UCEC pairs. (a-b) Scatter plots comparing the RC-exclusivity test with $N = 10^4$ permutations against (a) the R-exclusivity test and (b) the WR-exclusivity test (recursive) with weights $W^N_{RC}$. (c) The WR-exclusivity (recursive) $p$-values versus the WR-exclusivity (saddlepoint) $p$-values with weights $W^N_{RC}$. (d) Boxplots of the runtimes for computing the weighted test with the recusive formula (red) and with the saddlepoint approximation (blue) for each pair of genes in the datasets.}
  \label{fig:qqplots-and-runtime}
\end{figure*}

\begin{figure*}
  \centering
  \includegraphics[width=0.55\textwidth]{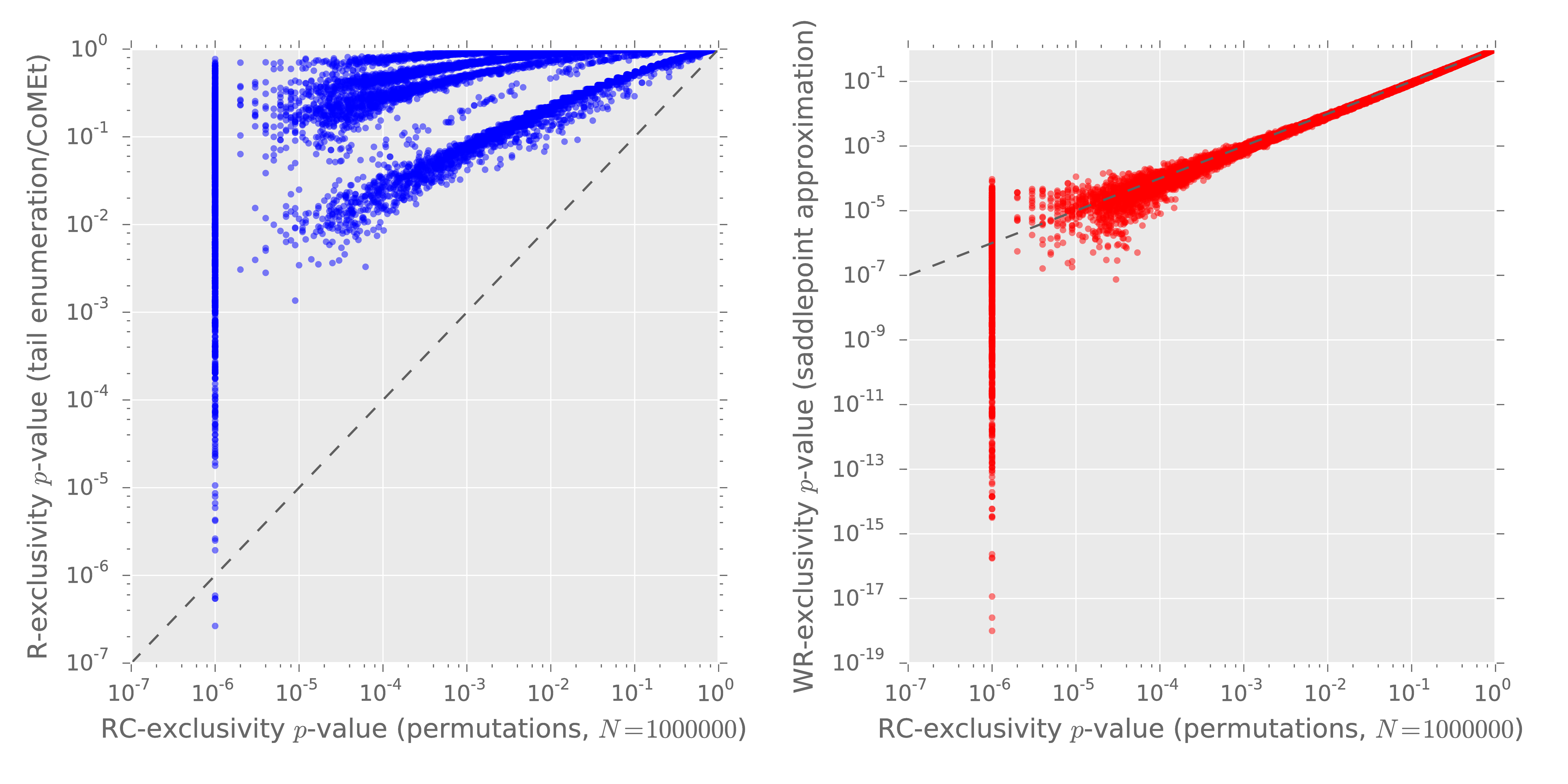}
  \caption{Comparison of the R-exclusivity test (left, $y$-axis) and the saddlepoint approximation of the WR-exclusivity test with weights $W^N_{RC}$ estimated from $N=10^3$ permutations (right, $y$-axis) to the R-exclusivity test ($x$-axis) on triples from the COADREAD dataset. The saddlepoint approximation uses weights $W^N_{RC}$ estimated from $N=10^3$ permutations, while the RC-exclusivity $p$-values were computed with $N = 10^6$ permutations.}
  \label{fig:WRE-vs-RCE}
\end{figure*}

\subsection{Approximating the RC-exclusivity test with the WR-exclusivity test}
\label{sec:approximation}
We compared the saddlepoint approximation of the WR-exclusivity test to the RC-exclusivity test using gene triples from the COADREAD dataset, again using the R-exclusivity test as a control. We computed $\PRC(M)$ with $N = 10^6$ permutations. We computed the saddlepoint approximation for $\PWR(M)$ using $W^N_{RC}$ with $N=10^3$ draws from $\SRC$, which is three orders of magnitude fewer than the number of permutations than we used to compute $\PRC(M)$. The $p$-values $\PR(M)$ and $\PRC(M)$ are weakly correlated in the tail ($\rho^2 = 0.67$ for $\PRC(M)<0.001$; see Figure~\ref{fig:WRE-vs-RCE}a). In contrast, the $\PWR(M)$ (saddlepoint) $p$-values provide an accurate approximation of the $\PRC(M)$ $p$-values. The RC-exclusivity and WR-exclusivity $p$-values are highly correlated in the tail ($\rho^2=0.948$ for $\PRC(M)<0.001$; see Figure~\ref{fig:WRE-vs-RCE}b).  Moreover, $\PWR(M)$ is an accurate estimate of $\PRC(M)$ to within one or more digits for most triples and within an order of magnitude for all triples. Furthermore, despite the much smaller number of permutations used to generate $W^N_{RC}$, $\PWR(M)$ provides smaller $p$-values than $\PRC(M)$ with a greater number of significant predictions, and is much faster than the permutation test.

\subsection{Mutually exclusive mutations in thyroid carcinomas}
\label{sec:thca}
We computed the WR-exclusivity $p$-values for all triples of genes that were each mutated in at least 5 of the 402 thyroid carcinomas in the THCA dataset. The WR-exclusivity test identifies 48 triples with significantly exclusive mutations (FDR $<0.001$), while the R-exclusivity test identifies 38 triples (FDR $<0.001$).

The top 25 ranked triples by both tests are identical, which is not surprising since THCA samples have low mutation rates compared to most cancer types (see~\citet{Vogelstein2013} and Section~\ref{sec:data}). In addition, the $p$-values for the top ranked triples are all within a few orders of magnitude, demonstrating that the two tests are very similar on this dataset.

Supplementary Table S1 shows the top triples, which include many known thyroid cancer genes. The top five triples include seven genes, five of which are well-known cancer genes with known roles in thyroid cancer~\citep{TCGA_THCA}: \emph{BRAF}, \emph{HRAS}, \emph{NRAS}, \emph{EIF1AX}, and \emph{ATM}. The other two genes are \emph{BDP1} and \emph{TG}, both of which may play a role in cancer. \citet{Woiwode2008} describe a role for \emph{BDP1} in AKT signaling, which was also noted in TCGA thyroid publication~\citep{TCGA_THCA}), although \emph{BDP1} is greater than 11,000 amino acids in length, so it may also accumulate many passenger mutations. \emph{TG} is the thyroglobulin gene, and is used as a tumor marker in papillary thyroid carcinoma~\citep{3}, which is the same subtype of thyroid cancer analyzed in TCGA.

\subsection{Mutually exclusive mutations in colorectal cancers and endometrial carcinomas}
\label{sec:coadread-ucec}
We expect that the difference between the R-exclusivity and WR-exclusivity tests would be more pronounced on cancer types with higher and highly variable mutation rates.  Thus, we computed $p$-values on triples of genes from colorectal cancers (COADREAD) and endometrial carcinomas (UCEC). We find that the WR-exclusivity test predicts more biologically interesting triples than the R-exclusivity test. The WR-exclusivity test identifies 5,290 and 6,835 triples (many of which overlap) with significantly mutually exclusive mutations (FDR $<$ 0.001) in the 224 COADREAD and 248 UCEC samples, respectively. In contrast, the R-exclusivity test computes 4 and 130 triples, respectively, with significantly mutually exclusive mutations (FDR $<0.001$).

Compared to the R-exclusivity test results, the highest ranked triples by the WR-exclusivity test include fewer long genes that tend to accumulate random, passenger mutations -- especially in samples with high mutation rates (Tables~\ref{table:coadread-triples} and~\ref{table:ucec-triples}).

\begin{table*}
  \centering
  \footnotesize
  \begin{tabular}{cclccc}
    \toprule
    \textbf{$\PR$ rank} & \textbf{$\PWR$ rank} & \textbf{Triple $M$} & \textbf{$\PR(M)$} & \textbf{$\PWR(M)$} & \textbf{Hypermutator mutations} \\
    \midrule
	1 & 2 & ACVR2A, PIK3CA, TP53 & $2.65 \cdot 10^{-7}$ & $2.54 \cdot 10^{-18}$ & 31\\
	2 & 32 & \textbf{APC}, BRAF, PRDM2 & $5.44 \cdot 10^{-7}$ & $2.30 \cdot 10^{-13}$ & 33\\
	2 & 33 & \textbf{APC}, BRAF, \textbf{WDFY3} & $5.44 \cdot 10^{-7}$ & $2.43 \cdot 10^{-13}$ & 32\\
	4 & 3 & \textbf{ATM}, PIK3CA, TP53 & $5.87 \cdot 10^{-7}$ & $1.15 \cdot 10^{-17}$ & 24\\
	5 & 81 & \textbf{APC}, BRAF, \textbf{FAT2} & $1.93 \cdot 10^{-6}$ & $6.48 \cdot 10^{-12}$ & 35\\
	\midrule
	6 & 1 & BRAF, KRAS, NRAS & $2.50 \cdot 10^{-6}$ & $9.95 \cdot 10^{-19}$ & 26\\
	1 & 2 & ACVR2A, PIK3CA, TP53 & $2.65 \cdot 10^{-7}$ & $2.54 \cdot 10^{-18}$ & 31\\
	4 & 3 & \textbf{ATM}, PIK3CA, TP53 & $5.87 \cdot 10^{-7}$ & $1.15 \cdot 10^{-17}$ & 24\\
	10 & 4 & ARID1A, TGFBR2, TP53 & $5.89 \cdot 10^{-6}$ & $1.76 \cdot 10^{-16}$ & 29\\
	9 & 5 & ABCA12, TGFBR2, TP53 & $4.29 \cdot 10^{-6}$ & $1.83 \cdot 10^{-16}$ & 28 \\
	\bottomrule
  \end{tabular}
  \caption{Five most significant triples identified by the R-exclusivity (upper 5) and WR-exclusivity (lower 5) tests on the COADREAD dataset. Genes in bold are among 600 longest genes (at least 9,560 amino acids).
  }
  \label{table:coadread-triples}
\end{table*}

On COADREAD, the WR-exclusivity test identifies ten different genes in the five most significant triples (Table~\ref{table:coadread-triples}). Nine of these genes are well-known cancer genes -- \emph{BRAF}, \emph{KRAS}, \emph{NRAS}, \emph{ACV2RA}, \emph{PIK3CA}, \emph{TP53}, \emph{ATM}, \emph{TGFBR2}, and \emph{ARID1A} -- while the tenth gene (\emph{ABCA12}) is known to have an association with colorectal cancers~\citep{Hlavata2012}. The R-exclusivity test results are similar -- two of the top five triples identified by the WR-exclusivity test are in the top five triples identified by the R-exclusivity test -- but the R-exclusivity test does not identify \emph{ARID1A}, \emph{TGFBR2}, \emph{KRAS}, or \emph{NRAS}. Further, the three additional genes identified by the R-exclusivity -- \emph{APC}, \emph{FAT2}, and \emph{WDFY3} -- are all in the top 600 longest genes in the human genome (at least 9,560 amino acids).  While mutations in \emph{APC} are well-known to play a role in colorectal cancers, there is currently little evidence for the roles of \emph{FAT2} or \emph{WDFY3} in cancer, and it is likely that these long genes have accumulated many passenger mutations, particularly in hypermutated samples.  Also of note is the fact that the number of hypermutator samples that contain mutations in the top triples from the WR-exclusivity test are not appreciably different from the number of hypermutator samples that contain mutations in the top triples from the R-exclusivity test.  This demonstrates that the WR-exclusivity is not systematically excluding hypermutator samples from consideration, but rather weighting the contribution of these samples appropriately in evaluating the significance of mutual exclusivity.

\begin{table*}
  \centering
  \footnotesize
  \begin{tabular}{cclccc}
    \toprule
    \textbf{$\PR$ rank} & \textbf{$\PWR$ rank} & \textbf{Triple $M$} & \textbf{$\PR(M)$} & \textbf{$\PWR(M)$} & \textbf{Hypermutator mutations} \\
    \midrule
	1 & 20 & \textbf{CACNA1E}, PTEN, TP53 & $3.11 \cdot 10^{-12}$ & $5.71 \cdot 10^{-30}$ & 77\\
	2 & 21 & \textbf{LAMA2}, PTEN, TP53 & $4.13 \cdot 10^{-12}$ & $8.05 \cdot 10^{-30}$ & 77\\
	3 & 29 & PTEN, \textbf{RYR2}, TP53 & $4.60 \cdot 10^{-12}$ & $4.85 \cdot 10^{-29}$ & 78\\
	4 & 28 & \textbf{NBEA}, PTEN, TP53 & $8.40 \cdot 10^{-12}$ & $3.32 \cdot 10^{-29}$ & 76\\
	5 & 39 & \textbf{FAT4}, PTEN, TP53 & $1.23 \cdot 10^{-11}$ & $3.0 \cdot 10^{-28}$ & 75\\
	\midrule
	22 & 1 & CTNNB1, RPL22, TP53 & $2.11 \cdot 10^{-10}$ & $1.20 \cdot 10^{-41}$ & 47\\
	44 & 2 & CTNNB1, KRAS, TP53 & $3.05 \cdot 10^{-9}$ & $1.10 \cdot 10^{-37}$ & 48\\
	55 & 3 & CTNNB1, MLL4, TP53 & $4.26 \cdot 10^{-8}$ & $5.78 \cdot 10^{-36}$ & 42\\
	57 & 4 & CTCF, CTNNB1, TP53 & $4.84 \cdot 10^{-8}$ & $3.29 \cdot 10^{-35}$ & 43\\
	60 & 5 & CTNNB1, \textbf{RYR1}, TP53 & $1.14 \cdot 10^{-7}$ & $9.51 \cdot 10^{-35}$ & 40 \\
	\bottomrule
  \end{tabular}
  \caption{Five most significant triples identified by the R-exclusivity (top 5) and WR-exclusivity (bottom 5) tests on the UCEC dataset. Notation as in Table~\ref{table:coadread-triples}.}
  \label{table:ucec-triples}
\end{table*}

On UCEC, the differences between the R-exclusivity and WR-exclusivity tests are even more pronounced. The WR-exclusivity test identifies seven genes in the top five most significant triples (Table~\ref{table:ucec-triples}). These include six genes  with known roles in cancer -- \emph{CTTNB1}, \emph{TP53}, \emph{RPL22},  \emph{KRAS},  \emph{CTCF}, and \emph{MLL4} -- with only one gene -- \emph{RYR1} -- with likely spurious mutations. In contrast,  the top five triples ranked by the R-exclusivity test include \emph{PTEN}, and \emph{TP53} -- two well-known cancer genes -- but also five genes with no known role in cancer that are all longer than 11,000 amino acids: \emph{CACNA1E}, \emph{LAMA2}, \emph{RYR2}, \emph{NBEA}, and \emph{FAT4}. Further, none of the top five triples identified by the WR-exclusivity test are in the top twenty R-exclusivity triples.  Finally, the R-exclusivity triples include many more mutations in hypermutator samples (ranging from mutations in 75 to 78 of the 81 hypermutators, versus 40 to 47 for the WR-exclusivity triples).  This further demonstrates how the results of the R-exclusivity test are skewed by hypermutator samples, while the WR-exclusivity test incorporates the contribution of these samples appropriately in evaluating the significance of mutual exclusivity.

\section{Discussion}

We introduce a weighted exact test for the mutual exclusivity of mutations in cancer. We use this test to approximate the permutational test for exclusivity where the number of mutations in each event and each sample are fixed.  To do so, we estimate per-event, per-sample mutation probabilities directly from the permutational distribution. We derive a recursive formula and a saddlepoint approximation of the $p$-value of the weighted test for event sets of any size, and we demonstrate the accuracy and efficiency of the saddlepoint approximation on genome-scale mutation datasets. Together, these contributions allow us to overcome the significant computational challenge of finding highly significant sets of mutually exclusive mutations conditioned on both the observed number of mutations per event and per sample.

We then demonstrate the weighted test on three datasets with hundreds of samples from TCGA, including colorectal and endometrial cancers that have high variability in the number of mutations per sample. The weighted test identifies sets of mutually exclusive mutations including known cancer genes in each dataset, and its results include many fewer long genes and mutations in hypermutator samples than do the results of the generalization of Fisher's exact test from CoMEt~\citep{Leiserson2015}.

There are several avenues for improving analyses with the weighted test. First, while we restricted our study to non-synonymous SNVs and indels, one should also test mutual exclusivity between other types of aberrations, such as copy number aberrations and gene fusions. We searched for mutually exclusive mutations by enumerating sets containing the most mutated genes, but the weighted test could easily be used in existing algorithms for optimizing mutual exclusivity scores (e.g., the MCMC from~\citet{Vandin2012}, the greedy approach from~\citet{Babur2015}) or to search for multiple sets simultaneously (e.g., from~\citet{Leiserson2015}). We estimated the per-event, per-sample mutation probability weights directly from the permutational distribution, but we also anticipate alternative methods for setting the weights that incorporate different event or sample attributes, such as gene length, to further reduce the number of false positives.

The weighted test may be of broader interest beyond searching for mutually exclusive mutations, both in other areas of computational biology and other disciplines. For example, statistical tests of ``presence-absence'' matrices with fixed row and column sums are a common tool in ecology for looking at species-associations, but can be computationally prohibitive (e.g., \cite{Miklos2004}). The weighted exact test presented here may offer a fast, alternative approach for computing the significance of associations with high accuracy.

\section{Acknowledgments}
The authors would like to acknowledge \cite{Manescu2015} for inspiring this work, and Uri Keich for generously helping us run their code and providing comments on our manuscript. This work is supported by US National Institutes of Health (NIH) grants R01HG005690, R01HG007069 and R01CA180776 to B.J.R. M.D.M.L.\ is supported by NSF fellowship GRFP DGE 0228243. B.J.R. is supported by a Career Award at the Scientific Interface from the Burroughs Wellcome Fund, an Alfred P. Sloan Research Fellowship and an NSF CAREER Award (CCF-1053753).

\bibliographystyle{plain}
\bibliography{references}

\begin{thebibliography}{10}

\bibitem{1}
\url{https://doi.org/10.7303/syn1710680.4}.

\bibitem{2}
\url{http://gdac.broadinstitute.org/runs/stddata__2016_01_28/data/THCA/20160128/gdac.broadinstitute.org_THCA.Mutation_Packager_Calls.Level_3.2016012800.0.0.tar.gz}.

\bibitem{3}
\url{http://www.mayomedicallaboratories.com/test-catalog/Clinical+and+Interpretive/62800}.

\bibitem{Babur2015}
{\"O}zg{\"u}n Babur, Mithat G{\"o}nen, B{\"u}lent~Arman Aksoy, Nikolaus
  Schultz, Giovanni Ciriello, et~al.
\newblock {Systematic identification of cancer driving signaling pathways based
  on mutual exclusivity of genomic alterations.}
\newblock {\em Genome biology}, 16:45, 2015.

\bibitem{Benjamini1995}
Y~Benjamini and Y~Hochberg.
\newblock {Controlling the false discovery rate: a practical and powerful
  approach to multiple testing}.
\newblock {\em Journal of the Royal Statistical Society Series B
  (Methodological)}, 57:289--300, 1995.

\bibitem{Butler2007}
Ronald~W Butler.
\newblock {\em Saddlepoint approximations with applications}, volume~22.
\newblock Cambridge University Press, 2007.

\bibitem{Ciriello2012}
Giovanni Ciriello, Ethan Cerami, Chris Sander, and Nikolaus Schultz.
\newblock {Mutual exclusivity analysis identifies oncogenic network modules.}
\newblock {\em Genome research}, 22(2):398--406, 2012.

\bibitem{Constantinescu2015}
Simona Constantinescu, Ewa Szczurek, Pejman Mohammadi, J{\"o}rg
  Rahnenf{\"u}hrer, and Niko Beerenwinkel.
\newblock {TiMEx: a waiting time model for mutually exclusive cancer
  alterations.}
\newblock {\em Bioinformatics (Oxford, England)}, 2015.

\bibitem{Gobbi2014}
Andrea Gobbi, Francesco Iorio, Kevin~J Dawson, David~C Wedge, David Tamborero,
  et~al.
\newblock {Fast randomization of large genomic datasets while preserving
  alteration counts.}
\newblock {\em Bioinformatics (Oxford, England)}, 30(17):i617--23, 2014.

\bibitem{Hanahan2011}
Douglas Hanahan and Robert~A Weinberg.
\newblock {Hallmarks of cancer: the next generation.}
\newblock {\em Cell}, 144(5):646--674, 2011.

\bibitem{Hlavata2012}
I~Hlavata, B~Mohelnikova-Duchonova, R~Vaclavikova, V~Liska, P~Pitule, P~Novak,
  et~al.
\newblock {The role of ABC transporters in progression and clinical outcome of
  colorectal cancer.}
\newblock {\em Mutagenesis}, 27(2):187--196, 2012.

\bibitem{Hong2013}
Yili Hong.
\newblock {On computing the distribution function for the Poisson binomial
  distribution.}
\newblock {\em Computational Statistics {\&} Data Analysis}, pages 41--51,
  2013.

\bibitem{Kim2015}
Yoo-Ah Kim, Dong-Yeon Cho, Phuong Dao, and Teresa~M Przytycka.
\newblock {MEMCover: integrated analysis of mutual exclusivity and functional
  network reveals dysregulated pathways across multiple cancer types.}
\newblock {\em Bioinformatics (Oxford, England)}, 31(12):i284--92, 2015.

\bibitem{Kim2016}
Yoo-Ah Kim, Sanna Madan, and Teresa~M Przytycka.
\newblock Wesme: Uncovering mutual exclusivity of cancer drivers and beyond.
\newblock {\em Bioinformatics (Oxford, England)}, page To appear, 2016.

\bibitem{Lawrence2013}
Michael~S Lawrence, Petar Stojanov, Paz Polak, Gregory~V Kryukov, Kristian
  Cibulskis, et~al.
\newblock {Mutational heterogeneity in cancer and the search for new
  cancer-associated genes.}
\newblock {\em Nature}, 499(7457):214--218, 2013.

\bibitem{Leiserson2013}
Mark D~M Leiserson, Dima Blokh, Roded Sharan, and Benjamin~J Raphael.
\newblock {Simultaneous identification of multiple driver pathways in cancer.}
\newblock {\em PLoS Computational Biology}, 9(5):e1003054, 2013.

\bibitem{Leiserson2015b}
Mark D~M Leiserson, Fabio Vandin, Hsin-Ta Wu, Jason~R Dobson, Jonathan~V
  Eldridge, et~al.
\newblock {Pan-cancer network analysis identifies combinations of rare somatic
  mutations across pathways and protein complexes.}
\newblock {\em Nature genetics}, 47(2):106--114, 2015.

\bibitem{Leiserson2015}
Mark D~M Leiserson, Hsin-Ta Wu, Fabio Vandin, and Benjamin~J Raphael.
\newblock {CoMEt: a statistical approach to identify combinations of mutually
  exclusive alterations in cancer.}
\newblock {\em Genome biology}, 16(1):160, 2015.

\bibitem{TCGA_AML}
Timothy~J Ley, Christopher Miller, Li~Ding, Benjamin~J Raphael, Andrew~J
  Mungall, et~al.
\newblock {Genomic and Epigenomic Landscapes of Adult De Novo Acute Myeloid
  Leukemia}.
\newblock {\em NEJM}, 368(22):2059--2074, 2013.

\bibitem{Manescu2015}
David Manescu and Uri Keich.
\newblock {A Symmetric Length-Aware Enrichment Test.}
\newblock {\em RECOMB}, pages 224--242, 2015.

\bibitem{Miklos2004}
I~Mikl{\'o}s and J~Podani.
\newblock {Randomization of presence-absence matrices: comments and new
  algorithms}.
\newblock {\em Ecology}, 85(1):86--92, 2004.

\bibitem{Miller2011}
Christopher~A Miller, Stephen~H Settle, Erik~P Sulman, Kenneth~D Aldape, and
  Aleksandar Milosavljevic.
\newblock {Discovering functional modules by identifying recurrent and mutually
  exclusive mutational patterns in tumors.}
\newblock {\em BMC medical genomics}, 4:34, 2011.

\bibitem{Milo2003}
Ron Milo, Nadav Kashtan, Shalev Itzkovitz, Mark~EJ Newman, and Uri Alon.
\newblock On the uniform generation of random graphs with prescribed degree
  sequences.
\newblock {\em arXiv preprint cond-mat/0312028}, 2003.

\bibitem{Mootha2003}
Vamsi~K Mootha, Cecilia~M Lindgren, Karl-Fredrik Eriksson, Aravind Subramanian,
  Smita Sihag, et~al.
\newblock {PGC-1alpha-responsive genes involved in oxidative phosphorylation
  are coordinately downregulated in human diabetes.}
\newblock {\em Nature genetics}, 34(3):267--273, 2003.

\bibitem{Roberts2014}
Steven~A Roberts and Dmitry~A Gordenin.
\newblock Hypermutation in human cancer genomes: footprints and mechanisms.
\newblock {\em Nature Reviews Cancer}, 14(12):786--800, 2014.

\bibitem{Ruffalo2015}
Matthew Ruffalo, Mehmet Koyut{\"u}rk, and Roded Sharan.
\newblock {Network-Based Integration of Disparate Omic Data To Identify "Silent
  Players" in Cancer.}
\newblock {\em PLoS computational biology}, 11(12):e1004595, 2015.

\bibitem{Subramanian2005}
Aravind Subramanian, Pablo Tamayo, Vamsi~K Mootha, Sayan Mukherjee, Benjamin~L
  Ebert, et~al.
\newblock {Gene set enrichment analysis: a knowledge-based approach for
  interpreting genome-wide expression profiles.}
\newblock {\em Proceedings of the National Academy of Sciences of the United
  States of America}, 102(43):15545--15550, 2005.

\bibitem{Szczurek2014}
Ewa Szczurek and Niko Beerenwinkel.
\newblock {Modeling mutual exclusivity of cancer mutations.}
\newblock {\em PLoS computational biology}, 10(3):e1003503, 2014.

\bibitem{TCGA_COADREAD}
{The Cancer Genome Atlas Research Network}.
\newblock {Comprehensive molecular characterization of human colon and rectal
  cancer.}
\newblock {\em Nature}, 487(7407):330--337, 2012.

\bibitem{TCGA_THCA}
{The Cancer Genome Atlas Research Network}.
\newblock {Integrated genomic characterization of papillary thyroid carcinoma.}
\newblock {\em Cell}, 159(3):676--690, 2014.

\bibitem{TCGA_UCEC}
{The Cancer Genome Atlas Research Network}, Cyriac Kandoth, Nikolaus Schultz,
  Andrew~D Cherniack, Rehan Akbani, et~al.
\newblock {Integrated genomic characterization of endometrial carcinoma.}
\newblock {\em Nature}, 497(7447):67--73, 2013.

\bibitem{Thomas2007}
Roman~K Thomas, Alissa~C Baker, Ralph~M Debiasi, Wendy Winckler, Thomas
  Laframboise, et~al.
\newblock {High-throughput oncogene mutation profiling in human cancer.}
\newblock {\em Nature genetics}, 39(3):347--351, 2007.

\bibitem{Vandin2011}
Fabio Vandin, Eli Upfal, and Benjamin~J Raphael.
\newblock {Algorithms for detecting significantly mutated pathways in cancer.}
\newblock {\em Journal of computational biology : a journal of computational
  molecular cell biology}, 18(3):507--522, 2011.

\bibitem{Vandin2012}
Fabio Vandin, Eli Upfal, and Benjamin~J Raphael.
\newblock {De novo discovery of mutated driver pathways in cancer.}
\newblock {\em Genome research}, 22(2):375--385, 2012.

\bibitem{Vogelstein2013}
Bert Vogelstein, Nickolas Papadopoulos, Victor~E Velculescu, Shibin Zhou,
  Luis~A Diaz, et~al.
\newblock {Cancer genome landscapes.}
\newblock {\em Science}, 339(6127):1546--1558, 2013.

\bibitem{Weinstein2013}
John~N Weinstein, Eric~A Collisson, Gordon~B Mills, Kenna R~Mills Shaw, Brad~A
  Ozenberger, et~al.
\newblock {The Cancer Genome Atlas Pan-Cancer analysis project}.
\newblock {\em Nature Genetics}, 45(10):1113--1120, 2013.

\bibitem{Wendl2011}
Michael~C Wendl, John~W Wallis, Ling Lin, Cyriac Kandoth, Elaine~R Mardis,
  et~al.
\newblock {PathScan: a tool for discerning mutational significance in groups of
  putative cancer genes.}
\newblock {\em Bioinformatics (Oxford, England)}, 27(12):1595--1602, 2011.

\bibitem{Woiwode2008}
Annette Woiwode, Sandra A~S Johnson, Shuping Zhong, Cheng Zhang, Robert~G
  Roeder, et~al.
\newblock {PTEN represses RNA polymerase III-dependent transcription by
  targeting the TFIIIB complex.}
\newblock {\em Molecular and cellular biology}, 28(12):4204--4214, 2008.

\bibitem{Yeang2008}
Chen-Hsiang Yeang, Frank McCormick, and Arnold Levine.
\newblock {Combinatorial patterns of somatic gene mutations in cancer.}
\newblock {\em FASEB journal : official publication of the Federation of
  American Societies for Experimental Biology}, 22(8):2605--2622, 2008.

\bibitem{Zhang2014}
Junhua Zhang, Ling-Yun Wu, Xiang-Sun Zhang, and Shihua Zhang.
\newblock {Discovery of co-occurring driver pathways in cancer.}
\newblock {\em BMC bioinformatics}, 15:271, 2014.

\end{thebibliography}
\includepdf[pages=-]{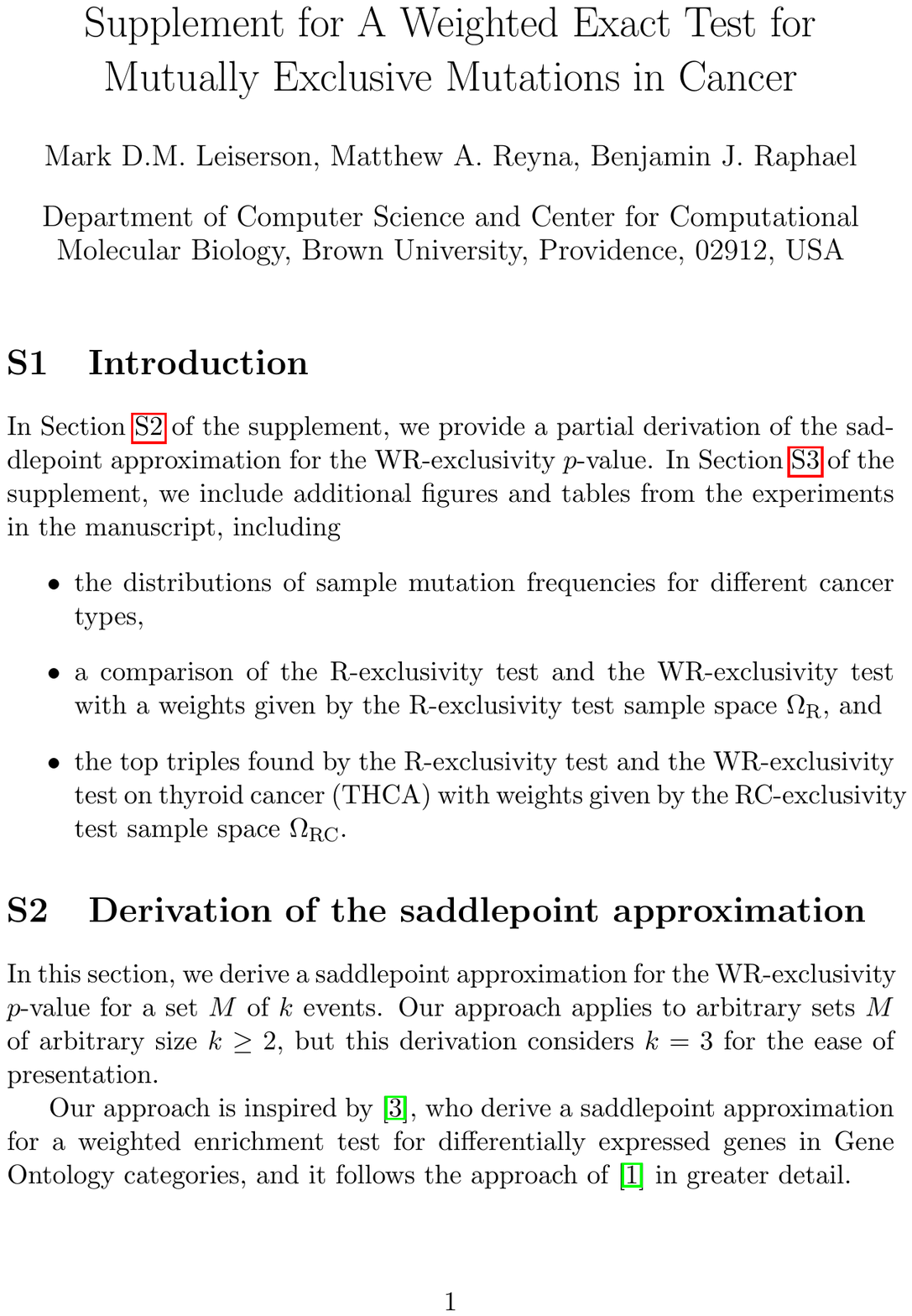}
\end{document}